\pdfoutput=1

\documentclass{aastex631}
\usepackage{amsmath,bm}
\usepackage{graphicx}
\usepackage{mathrsfs}
\usepackage{url}
\usepackage{CJKutf8}
\usepackage{ulem}
\usepackage{wasysym}
\usepackage{booktabs}
\usepackage{multirow}

\received{2022}
\revised{\today}
\accepted{2022}

\shorttitle{331P}
\shortauthors{Hui \& Jewitt 2022}

\begin{document}

\title{
Fragment Dynamics in Active Asteroid 331P/Gibbs
}

\correspondingauthor{Man-To Hui}
\email{mthui@must.edu.mo}

\author{\begin{CJK}{UTF8}{bsmi}Man-To Hui (許文韜)\end{CJK}}
\affiliation{State Key Laboratory of Lunar and Planetary Science, 
Macau University of Science and Technology, 
Avenida Wai Long, Taipa, Macau}

\author{David Jewitt}
\affiliation{Department of Earth, Planetary and Space Sciences, UCLA, 
595 Charles Young Drive East, Los Angeles, CA 90095-1567, USA}


\begin{abstract}

We present a dynamical analysis of the fragmented active asteroid 331P/Gibbs. Using archival images taken by the Hubble Space Telescope from 2015 to 2018, we measured the astrometry of the primary and the three brightest (presumably the largest) components. Conventional orbit determination revealed a high-degree of orbital similarity between the components. We then applied a fragmentation model to fit the astrometry,  obtaining  key parameters including the fragmentation epochs and separation velocities. Our best-fit models show that Fragment B separated from the primary body at a speed of $\sim$1 cm s$^{-1}$ between 2011 April and May, whereas two plausible scenarios were identified for Fragments A and C. The former split either from the primary or from Fragment B, in 2011 mid-June at a speed of $\sim$8 cm s$^{-1}$, and the latter split from Fragment B either in late 2011 or between late 2013 and early 2014, at a speed of $\sim$0.7-0.8 cm s$^{-1}$. The results are  consistent with rotational disruption as the mechanism causing the cascading fragmentation of the asteroid, as suggested by the rapid rotation of the primary. The fragments constitute the youngest known  asteroid cluster, providing us with a great opportunity to study asteroid fragmentation and formation of asteroid clusters.

\end{abstract}


\section{Introduction}
\label{sec_intro}

Active asteroids are a newly discovered class of small solar system bodies having dual characteristics of both comets and asteroids. While their orbits are dynamically asteroidal \citep[Jupiter Tisserand invariant $T_{\rm J} \ge 3.08$;][]{2015aste.book..221J}, they are observed to show cometary features visually similar to ordinary comets. The existence of active asteroids challenges the traditional  view of comets and asteroids as two distinct classes of small solar system body. Since the discovery of the prototype active asteroid 133P/Elst-Pizarro \citep{2006Sci...312..561H}, more than two dozen active asteroids have been identified \citep{2022arXiv220301397J}. However, while their cometary morphology unequivocally indicates mass loss, several different physical mechanisms are responsible for the observed activity, including the sublimation of water ice, impact, rotational disruption, and thermal fracturing \citep[e.g.,][]{2022arXiv220301397J}. The enigmatic population still remains poorly understood to date.

Active asteroid 331P/Gibbs was discovered in 2012 \citep{2012CBET.3069....1G}. Its orbit lies in the outer main belt, with  semimajor axis $a = 3.01$ au, eccentricity $e = 0.04$, and inclination $i = 9\fdg7$, corresponding to a Jupiter Tisserand invariant of $T_{\rm J} = 3.23$. The  observed mass loss from  331P was at first interpreted to result from an impulsive event, consistent with an impact origin, estimated to have occurred in 2011 July \citep{2012ApJ...761L..12M,2012ApJ...759..142S}. Deep imaging with the Keck telescope in 2014 revealed four fragments embedded in a diffuse particle trail, in addition to the kilometer-sized primary nucleus, the latter with a rotation period $3.26 \pm 0.01$ hour  \citep{2015ApJ...802L...8D}.   Deeper imaging with the Hubble Space Telescope (HST) from 2015 to 2018, revealed a wealth of detail, including 19 fragments with radii in the range  0.04 km to 0.11 km (geometric albedo 0.05 assumed) one of which (331P-A) is a likely contact binary \citep{2021AJ....162..268J}.  The fragments follow a size distribution with a differential power-law index of $\gamma = 3.7\pm 0.1$ to $4.1\pm 0.1$,  have a combined mass approximately 1\% of that of the primary and ejection velocities  crudely estimated as $\sim$10 cm s$^{-1}$ \citep{2021AJ....162..268J}. Intriguingly, 331P is also a member of a dynamical family with a separation age of $1.5 \pm 0.1$ Myr, but no other active asteroids from the young cluster have been found \citep{2014Icar..231..300N}. The relationship between the current mass loss from 331P and  the catastrophic event which gave birth to the asteroid cluster, if any, remains unclear.

In this paper, we analyze the same archival HST observations as in \cite{2021AJ....162..268J} but with the prime objective being to study the dynamics of the fragments. We detail the observations and data reduction in Section \ref{sec_obs}, present the astrometric and dynamical analyses in Section \ref{sec_nls}, the results  in Section \ref{sec_dsc} and a summary in Section \ref{sec_sum}.

\begin{figure*}
\epsscale{1.0}
\begin{center}
\plotone{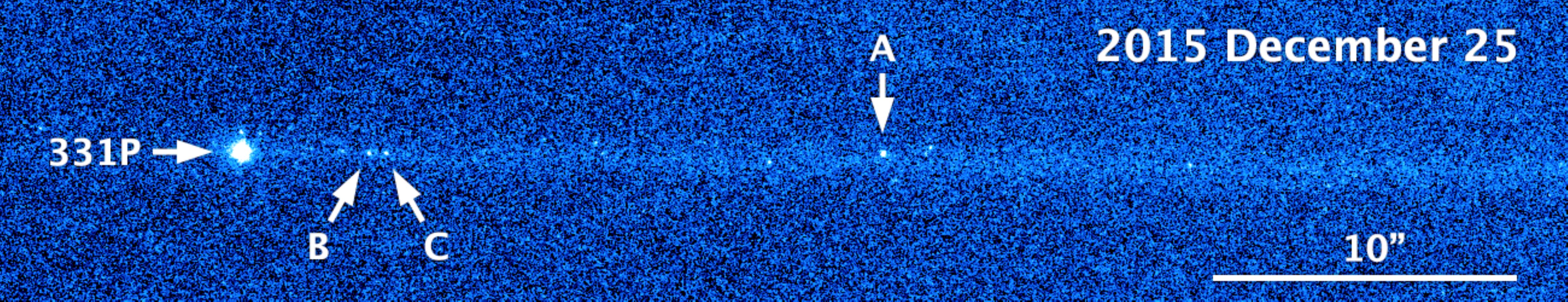}
\caption{
Identification of the primary body, 331P, and the three brightest fragments 331P-A, -B and -C, analysed in this work.  A dust trail is evident spanning the image and other fragments are visible in the figure but are not studied here.  The region shown is $\sim\!7.5 \times 10^4$ km in width projected to the distance of the object. J2000 equatorial north is up and east is left. A scale bar of 10\arcsec~in length is given on the lower right corner.
\label{fig:obs}
} 
\end{center} 
\end{figure*}

\begin{deluxetable*}{lccccccccc}
\tablecaption{Observing Geometry of Active Asteroid 331P/Gibbs
\label{tab_obs}}
\tablewidth{0pt}
\tablehead{
\colhead{Date} & \colhead{\# Images} &  \colhead{$t_{\rm exp}$ (s)\tablenotemark{a}} & \colhead{$r_H$ (au)\tablenotemark{b}} & \colhead{${\it \Delta}$ (au)\tablenotemark{c}} & \colhead{$\alpha$ (\degr)\tablenotemark{d}} & \colhead{$\nu$ (\degr)\tablenotemark{e}} & \colhead{$\theta_{-\odot}$ (\degr)\tablenotemark{f}} & \colhead{$\theta_{-\bf v}$ (\degr)\tablenotemark{g}} & \colhead{$\psi$ (\degr)\tablenotemark{h}}
}
\startdata
2015 Dec 25 & 24 & 368 & 2.906 & 2.011 & 9.7 & 39.3 & 64.7 & 269.0 & 3.9 \\
2015 Dec 28 & 12 & 368 & 2.907 & 2.030 & 10.5 & 39.8 & 66.2 & 269.0 & 4.0 \\
2016 Feb 13 & 24 & 368 & 2.920 & 2.575 & 19.4 & 49.3 & 78.3 & 269.6 & 3.6 \\
2017 Feb 13 & 24 & 438 & 3.059 & 2.111 & 6.2 & 118.8 & 331.1 & 286.2 & 4.3 \\
2017 Mar 8 & 25 & 438 & 3.068 & 2.109 & 5.8 & 123.0 & 70.4 & 286.7 & 3.4 \\
2018 May 17 & 10 & 438 & 3.122 & 2.113 & 1.6 & 197.4 & 169.0 & 274.5 & -1.5 \\
2018 May 26 & 20 & 438 & 3.121 & 2.126 & 4.3 & 198.9 & 125.4 & 274.9 & -2.2 \\
2018 Jul 3 & 20 & 453 & 3.115 & 2.402 & 15.2 & 205.3 & 110.0 & 275.2 & -3.7 \\
\enddata
\tablenotetext{a}{Individual median exposure time.}
\tablenotetext{b}{Heliocentric distance.}
\tablenotetext{c}{HST-centric distance.}
\tablenotetext{d}{Phase angle (Sun-331P-HST).}
\tablenotetext{e}{True anomaly.}
\tablenotetext{f}{Position angle of the projected antisolar direction.}
\tablenotetext{g}{Position angle of the projected negative heliocentric velocity of 331P.}
\tablenotetext{h}{Orbital plane angle (between HST and orbital plane of 331P).}
\end{deluxetable*}

\section{Observations \& Data Reduction}
\label{sec_obs}

We collected archival HST observations of active asteroid 331P taken under General Observer programs 14192, 14475, 14798, and 15360 (PI: M. Drahus) using the HST Data Search\footnote{\url{https://archive.stsci.edu/hst/search.php}} in the Mikulski Archive for Space Telescopes. The target was imaged from eight different epochs by HST's Wide Field Camera 3 (WFC3), which houses two 2k $\times$ 4k pixel CCDs separated by a 31-pixel gap and has an image scale of 0\farcs04 pixel$^{-1}$ in the UVIS channel \citep{2022wfci.book...14D}. The long-pass F350LP filter, having effective wavelength 5846 \AA, FWHM 4758 \AA~and maximum throughput $\sim$29\%, was used for all of the images. Because the telescope was tracked at the nonsidereal rate of 331P, background sources are clearly trailed, and even visibly curved in some of the images, due to the parallactic motion of HST around the Earth. The observing geometry of 331P is given in Table \ref{tab_obs}.

We started with the calibrated images including charge transfer efficiency correction but no geometric distortion correction. These images were plagued with abundant hot pixels and random cosmic ray hits. While the hot pixels were flagged and masked in accordance with the data quality arrays accompanying the science images, we applied the {\tt L.A. Cosmic} algorithm \citep{2001PASP..113.1420V} to remove cosmic ray strikes, thereby obtaining much cleaner resulting images. We checked that the pixel counts of the components of 331P and neighbouring unsaturated background sources remained untouched by the cosmic ray removal algorithm. Next,   {\tt AstroDrizzle} \citep{2012drzp.book.....G} was used to correct the cleaned images for geometric distortion. In order to further suppress remaining cosmic ray artefacts and to improve the signal-to-noise ratio (S/N) of the target, images from the same visits were median combined with alignment on the primary nucleus of 331P. Images in which the active asteroid was strongly influenced or obscured by saturated bright sources were discarded. As a result, most  background star trails were effectively removed in the combined stacks. 

The combined HST images of 331P have a range of sensitivities, both because of the changing viewing geometry and because different numbers of orbits and total observing times were secured in different visits (Table \ref{tab_obs}).  While a total of 19 fragments, labeled 331P-A through 331P-S, are clearly identified in the data \citep{2021AJ....162..268J}, only the  brightest three (Figure \ref{fig:obs}) can be identified in images taken across the full interval from 2015 December to 2018 July. For this reason, the following analysis applies only to these three fragments, 331P-A, -B and -C, in addition to the primary body, for which the astrometric data are of sufficient number and quality to permit a useful dynamical analysis.

\section{Analysis}
\label{sec_nls}

\subsection{Astrometry}
\label{ssec_astrom}

The HST/WFC3 images only contained coarse world coordinate system (WCS) solutions with guide stars in their headers, and therefore updating the solutions using field stars would be necessary before we could determine best-fit orbital solutions for the primary nucleus of 331P and its fragments. Because stars were all trailed significantly in individual HST/WFC3 images, traditional centroiding algorithms which treat stars as point sources could not be used. Instead, we adopted a technique specifically suited for trailed images, in which a source is modelled as a trapezoid in the along-track direction and a Gaussian in the cross-track direction. For each trail, the peak value and pixel coordinates of the centroid, along with the trail length, width, and orientation were free parameters to be fitted. The local sky background was determined using an annulus centered around the corresponding trail, with the inner and outer ranges respectively twice and four times the dimensions of the trail aperture. For each visit, we picked the image in which the trails of background sources appear to be the most straight by visual inspection, so as to better avoid introduction of a systematic bias due to curvature being unaccounted for in the trail model. We then updated the WCS solutions by a linear fit of the pixel coordinates to the Gaia Data Release 2 catalogue \citep{2018yCat.1345....0G}.

Measuring the pixel coordinates of the primary and the brightest fragments of 331P was straightforward in the median-combined images, as each of them was simply treated as a point source to be least-square fitted. Thereby, with the updated WCS solutions, we were able to convert the best-fit pixel coordinates of the primary and components of 331P in the corresponding HST visits to J2000 equatorial coordinates in terms of R.A. and decl. The associated uncertainties were obtained through propagation of the centroid errors and the counterparts in astrometric solutions, with the latter being dominant. We tabulate the astrometry of the components of 331P along with the corresponding uncertainties in Table \ref{tab_astrom}.

\begin{deluxetable*}{c@{\extracolsep{\fill}}c@{\extracolsep{\fill}}c@{\extracolsep{\fill}}c@{\extracolsep{\fill}}c@{\extracolsep{\fill}}c@{\extracolsep{\fill}}c@{\extracolsep{\fill}}c@{\extracolsep{\fill}}c@{\extracolsep{\fill}}c@{\extracolsep{\fill}}c@{\extracolsep{\fill}}c@{\extracolsep{\fill}}c}
\rotate
\tablewidth{0pt}
\tabletypesize{\scriptsize}
\tablecaption{Astrometry of the Primary and Components A, B, \& C of Active Asteroid 331P/Gibbs
\label{tab_astrom}}
\tablehead{
\colhead{Time} & \multicolumn{3}{c}{Primary} &  \multicolumn{3}{c}{Fragment A} & \multicolumn{3}{c}{Fragment B} & \multicolumn{3}{c}{Fragment C} \\ \cmidrule(lr){2-4} \cmidrule(lr){5-7} \cmidrule(lr){8-10} \cmidrule(lr){11-13}
\colhead{(UTC)} & 
\colhead{R.A. (h m s)} & \colhead{Decl. (\degr~\arcmin~\arcsec)} & \colhead{Error (\arcsec)} & 
\colhead{R.A. (h m s) } & \colhead{Decl. (\degr~\arcmin~\arcsec)} & \colhead{Error (\arcsec)} & 
\colhead{R.A. (h m s)} & \colhead{Decl. (\degr~\arcmin~\arcsec)} & \colhead{Error (\arcsec)} & 
\colhead{R.A. (h m s)} & \colhead{Decl. (\degr~\arcmin~\arcsec)} & \colhead{Error (\arcsec)}
}
\startdata
2015 Dec 25.428287 & 
04 16 31.2730 & +12 38 01.611 & 0.035 0.006 & 
04 16 29.8323 & +12 38 01.542 & 0.031 0.010 &
04 16 30.9851 & +12 38 01.541 & 0.044 0.029 &
04 16 30.9475 & +12 38 01.567 & 0.039 0.021 \\
2015 Dec 25.492963 & 
04 16 28.9437 & +12 37 52.882 & 0.094 0.040 &
04 16 27.5046 & +12 37 52.813 & 0.076 0.040 &
04 16 28.6576 & +12 37 52.819 & 0.092 0.040 &
04 16 28.6181 & +12 37 52.817 & 0.092 0.040 \\
2015 Dec 25.559225 & 
04 16 26.5694 & +12 37 44.177 & 0.040 0.040 &
04 16 25.1291 & +12 37 44.100 & 0.040 0.040 &
04 16 26.2827 & +12 37 44.114 & 0.040 0.040 &
04 16 26.2440 & +12 37 44.129 & 0.080 0.080 \\
2015 Dec 25.625486 & 
04 16 24.1958 & +12 37 35.506 & 0.040 0.040 &
04 16 22.7556 & +12 37 35.424 & 0.040 0.040 &
04 16 23.9109 & +12 37 35.442 & 0.040 0.040 &
04 16 23.8717 & +12 37 35.456 & 0.040 0.040 \\
2015 Dec 28.077894 & 
04 15 00.0126 & +12 32 36.090 & 0.006 0.004 &
04 14 58.5905 & +12 32 36.004 & 0.009 0.008 &
04 14 59.7303 & +12 32 36.029 & 0.033 0.034 &
04 14 59.6912 & +12 32 36.044 & 0.012 0.012 \\
2015 Dec 28.143310 & 
04 14 57.8805 & +12 32 28.588 & 0.200 0.200 &
04 14 56.4571 & +12 32 28.511 & 0.200 0.200 &
04 14 57.5975 & +12 32 28.552 & 0.200 0.200 &
04 14 57.5584 & +12 32 28.531 & 0.200 0.200 \\
2016 Feb 13.159294 & 
04 15 22.8222 & +13 00 32.481 & 0.040 0.040 &
04 15 21.7378 & +13 00 32.519 & 0.040 0.040 &
&&&
&& \\
2016 Feb 13.236181 & 
04 15 25.1671 & +13 00 42.757 & 0.040 0.040 &
04 15 24.0830 & +13 00 42.810 & 0.040 0.040 &
&&&
&& \\
2016 Feb 13.302442 & 
04 15 27.3454 & +13 00 52.596 & 0.040 0.040 &
04 15 26.2611 & +13 00 52.649 & 0.080 0.080 &
&&&
&& \\
2016 Feb 13.368692 & 
04 15 29.5288 & +13 01 02.439 & 0.040 0.040 &
04 15 28.4451 & +13 01 02.491 & 0.040 0.040 &
&&&
&& \\
2017 Feb 13.098513 & 
10 25 20.0925 & -03 38 30.866 & 0.020 0.019 &
10 25 18.8912 & -03 38 25.800 & 0.021 0.020 &
10 25 19.8566 & -03 38 29.849 & 0.023 0.023 &
10 25 19.8093 & -03 38 29.632 & 0.021 0.020 \\
2017 Feb 13.170463 & 
10 25 16.8981 & -03 38 13.799 & 0.010 0.011 &
10 25 15.6964 & -03 38 08.716 & 0.009 0.010 &
10 25 16.6626 & -03 38 12.783 & 0.014 0.017 &
10 25 16.6149 & -03 38 12.561 & 0.014 0.015 \\
2017 Feb 13.236690 & 
10 25 14.0273 & -03 37 56.884 & 0.007 0.012 &
10 25 12.8255 & -03 37 51.812 & 0.008 0.013 &
10 25 13.7922 & -03 37 55.863 & 0.014 0.016 &
10 25 13.7441 & -03 37 55.639 & 0.019 0.023 \\
2017 Feb 13.296377 & 
10 25 11.5174 & -03 37 40.210 & 0.010 0.026 &
10 25 10.3162 & -03 37 35.125 & 0.011 0.024 &
10 25 11.2829 & -03 37 39.216 & 0.015 0.030 &
10 25 11.2341 & -03 37 38.970 & 0.013 0.026 \\
2017 Feb 13.362604 & 
10 25 08.6441 & -03 37 23.210 & 0.009 0.020 &
10 25 07.4424 & -03 37 18.131 & 0.008 0.018 &
10 25 08.4079 & -03 37 22.194 & 0.029 0.034 &
10 25 08.3607 & -03 37 21.968 & 0.011 0.021 \\
2017 Mar 08.677274 & 
10 07 59.2848 & -01 25 11.466 & 0.012 0.019 &
10 07 58.0890 & -01 25 06.242 & 0.011 0.017 &
10 07 59.0500 & -01 25 10.381 & 0.051 0.058 &
10 07 59.0019 & -01 25 10.208 & 0.012 0.019 \\
2017 Mar 08.768877 & 
10 07 55.0979 & -01 24 31.750 & 0.200 0.200 &
10 07 53.9018 & -01 24 26.524 & 0.200 0.200 &
10 07 54.8642 & -01 24 30.687 & 0.200 0.200 &
10 07 54.8155 & -01 24 30.477 & 0.200 0.200 \\
2017 Mar 08.835093 & 
10 07 52.4294 & -01 24 05.641 & 0.200 0.200 &
10 07 51.2334 & -01 24 00.415 & 0.200 0.200 &
&&&
&& \\
2017 Mar 08.901319 & 
10 07 49.7653 & -01 23 39.545 & 0.200 0.200 &
10 07 48.5700 & -01 23 34.326 & 0.200 0.200 &
10 07 49.5312 & -01 23 38.515 & 0.200 0.200 &
10 07 49.4836 & -01 23 38.278 & 0.200 0.200 \\
2017 Mar 08.967535 & 
10 07 47.1056 & -01 23 13.436 & 0.200 0.200 &
10 07 45.9096 & -01 23 08.215 & 0.200 0.200 &
10 07 46.8725 & -01 23 12.376 & 0.200 0.200 &
10 07 46.8233 & -01 23 12.179 & 0.200 0.200 \\
2018 May 17.583999 & 
15 32 25.0055 & -14 27 20.002 & 0.009 0.006 &
15 32 23.7058 & -14 27 18.796 & 0.010 0.007 &
15 32 24.7090 & -14 27 19.666 & 0.014 0.016 &
15 32 24.6713 & -14 27 19.646 & 0.013 0.011 \\
2018 May 17.649618 & 
15 32 21.9177 & -14 26 58.985 & 0.040 0.040 &
15 32 20.6178 & -14 26 57.776 & 0.040 0.040 &
15 32 21.6199 & -14 26 58.679 & 0.040 0.040 &
15 32 21.5836 & -14 26 58.639 & 0.040 0.040 \\
2018 May 26.390353 & 
15 25 39.2484 & -13 41 50.450 & 0.006 0.005 &
15 25 37.9666 & -13 41 49.178 & 0.011 0.011 &
15 25 38.9545 & -13 41 50.127 & 0.017 0.015 &
15 25 38.9186 & -13 41 50.066 & 0.009 0.008 \\
2018 May 26.469062 & 
15 25 35.4895 & -13 41 29.011 & 0.040 0.040 &
15 25 34.2076 & -13 41 27.728 & 0.040 0.040 &
15 25 35.1950 & -13 41 28.682 & 0.040 0.040 &
15 25 35.1601 & -13 41 28.623 & 0.040 0.040 \\
2018 May 26.601238 & 
15 25 29.6242 & -13 40 49.975 & 0.026 0.004 &
15 25 28.3429 & -13 40 48.707 & 0.024 0.008 &
15 25 29.3306 & -13 40 49.622 & 0.027 0.012 &
15 25 29.2950 & -13 40 49.589 & 0.026 0.008 \\
2018 May 26.667674 & 
15 25 26.6726 & -13 40 30.450 & 0.018 0.004 &
15 25 25.3913 & -13 40 29.168 & 0.017 0.007 &
15 25 26.3789 & -13 40 30.114 & 0.022 0.018 &
15 25 26.3433 & -13 40 30.058 & 0.020 0.012 \\
2018 Jun 02.540538 & 
15 20 36.2912 & -13 08 28.232 & 0.040 0.040 &
15 20 35.0352 & -13 08 26.915 & 0.040 0.040 &
15 20 36.0034 & -13 08 27.893 & 0.040 0.040 &
15 20 35.9688 & -13 08 27.846 & 0.040 0.040 \\
2018 Jul 03.426128 & 
15 08 15.7534 & -11 44 20.592 & 0.100 0.100 &
15 08 14.6690 & -11 44 19.482 & 0.100 0.100 &
15 08 15.5041 & -11 44 20.329 & 0.100 0.100 &
15 08 15.4724 & -11 44 20.264 & 0.100 0.100 \\
2018 Jul 03.492251 & 
15 08 15.3940 & -11 44 16.960 & 0.100 0.100 &
15 08 14.3104 & -11 44 15.856 & 0.100 0.100 &
15 08 15.1452 & -11 44 16.675 & 0.100 0.100 &
15 08 15.1123 & -11 44 16.611 & 0.100 0.100 \\
2018 Jul 03.624751 & 
15 08 14.6926 & -11 44 09.790 & 0.100 0.100 &
15 08 13.6093 & -11 44 08.675 & 0.100 0.100 &
15 08 14.4447 & -11 44 09.512 & 0.100 0.100 &
15 08 14.4128 & -11 44 09.457 & 0.100 0.100 \\
2018 Jul 03.690874 & 
15 08 14.3511 & -11 44 06.246 & 0.100 0.100 &
15 08 13.2685 & -11 44 05.148 & 0.100 0.100 &
15 08 14.1026 & -11 44 05.956 & 0.100 0.100 &
15 08 14.0728 & -11 44 05.918 & 0.100 0.100 \\
\enddata
\tablecomments{The R.A. and decl. coordinates are referred to the J2000 equatorial system. The reported errors are $1\sigma$ formal uncertainties of the astrometric measurements, primarily dominated by systematic errors in the WCS solutions. Technically, the first columns of the errors are uncertainties in the east-west direction at the corresponding J2000 equatorial coordinates.}
\end{deluxetable*}

\subsection{Orbit Determination}
\label{ssec_orb}

\begin{deluxetable*}{lCcccc}
\tablecaption{Best-fit Orbital Solutions for the Primary and Fragments A, B, \& C of Active Asteroid 331P/Gibbs
\label{tab_orb}}
\tablewidth{0pt}
\tablehead{
\multicolumn{2}{c}{Quantity}  & 
\colhead{Primary} &
\colhead{Fragment A} & 
\colhead{Fragment B} &
\colhead{Fragment C}
}
\startdata
Eccentricity & e
       & 0.039874582(21)
       & 0.03987205(39)
       & 0.03987449(81)
       & 0.03987430(53) \\ 
Semimajor axis (au) & a
       & 3.007603337(39)
       & 3.0076255(26)
       & 3.0076105(54)
       & 3.0076092(35) \\
Perihelion distance (au) & q
       & 2.887676412(80)
       & 2.8877053(37)
       & 2.8876836(77)
       & 2.8876829(50) \\ 
Inclination (\degr) & i
       & 9.7434076(11)
       & 9.7434091(13)
       & 9.7434115(23)
       & 9.7434075(17) \\ 
Longitude of ascending node (\degr) & {\it \Omega}
                 & 216.8118502(58)
                 & 216.812121(13)
                 & 216.811851(23)
                 & 216.811859(16) \\ 
Argument of perihelion (\degr) & \omega
                 & 181.374530(26)
                 & 181.3851(13)
                 & 181.3777(27)
                 & 181.3768(17) \\ 
Time of perihelion (TDB) & t_{\rm p}
                  & 2020 Sep 18.61468(15)
                  & 2020 Sep 18.7065(84)
                  & 2020 Sep 18.641(18)
                  & 2020 Sep 18.636(11) \\
\hline
\multicolumn{2}{l}{Observed arc}
& 2004 Aug 26 - 2022 Jan 26
& 2015 Dec 25 - 2018 Jul 3
& 2015 Dec 25 - 2018 Jul 3
& 2015 Dec 25 - 2018 Jul 3 \\
\multicolumn{2}{l}{Number of observations used}
& 172
& 31
& 26
& 26 \\
\multicolumn{2}{l}{Normalized rms residual}
& 1.103
& 0.843
& 0.933
& 0.851 \\
\enddata
\tablecomments{The orbital elements are all referenced to the heliocentric J2000 ecliptic, all at a osculation epoch of TDB 2018 Jul 3.0, where TDB is Barycentric Dynamical Time. Here the uncertainties of the orbital elements are all $1\sigma$ formal errors.}
\end{deluxetable*}

In addition to our astrometric measurements, we downloaded astrometry of the primary from the Minor Planet Center Database Search,\footnote{\url{https://minorplanetcenter.net/db_search}} which covers a much longer observed arc. For our astrometric measurements, we assigned a weighting scheme in accordance with the corresponding measurement errors, whereas we had to debias and weight the counterpart from the Minor Planet Center following the methods described in \citet{2015Icar..245...94F} and \citet{2017Icar..296..139V}, respectively, because the astrometry was measured with different star catalogues and no information of their measurement errors is available. We then proceeded to find orbital solutions for the primary and the brightest components of 331P in {\tt FindOrb}, an N-body orbit determination software package written by B. Gray,\footnote{The package is available from \url{https://github.com/Bill-Gray/find_orb}.} which took into account  solar gravity, as well as perturbations from the eight major planets, Pluto, the Moon, and the 16 most massive asteroids in the main belt. Positions and masses of the perturbers were adopted from the planetary and lunar ephemeris DE440 \citep{2021AJ....161..105P}. The first post-Newtonian corrections were also implemented in the software package. In a preliminary test, we were aware that almost half of our astrometric measurements from the HST images for the primary had observed-minus-calculated (O-C) residuals more than three times larger than the reported measurement errors (no more than $\sim\!10$ mas), five of which even had residuals over ten times larger. By looking into these measurements more closely, we found that the major discrepancies all existed in the along-track direction. We attempted to refine the astrometric reduction for these images by slightly adjusting the initial guess values in fitting the star trails. However, no improvement beyond the noise level could be achieved. We then visually inspected the images, finding that the star trails appear more curved due to the parallactic motion of the HST orbiting around Earth. Therefore, we concluded that the O-C residuals of these measurements reflected the systematic errors and inflated the measurement errors accordingly. Astrometry of the primary downloaded from the Minor Planet Center with O-C residuals over three times the assigned uncertainties was simply discarded, because their O-C residuals can be as enormous as a few arcsec, orders of magnitude greater than those of our worst HST astrometry. We also inflated the measurement errors for the astrometry of the fragments from the same sets of images using the results we obtained for the primary. Consequently, the O-C residuals of the astrometry of the fragments are all consistent with the inflated errors, suggesting the validity of the modification. We summarize the best-fit orbital solutions in Table \ref{tab_orb}, from which the similarity of the orbits can be readily noticed.

\subsection{Fragmentation}
\label{ssec_frg}

The orbital similarities of the components of 331P are indicative of their common origin.  We are interested to know whether the fragments 331P-A, -B and -C were  released from the primary at one time, or staggered at different times.  In addition, we consider the possibility that some fragments may have been released  from other fragments (in so-called ``cascading fragmentation''), not directly from the 331P parent body.  We address these questions by carefully examining the pair-wise separations of the fragments both from the parent and from each other.  

We applied two different approaches to investigate the fragmentation of the active asteroid. First, we backtracked the orbits of the components by means of backward integrating the obtained best-fit orbits summarized in Table \ref{tab_orb}. In order to fully encompass the orbital uncertainties of the components, we applied the Cholesky decomposition method to generate 1000 Monte Carlo (MC) clones for each component based on the corresponding covariance matrices of the orbital elements. We did not attempt to compute and track the mutual distances between the clones of the fragments and those of the primary, because we only used results from this approach to guide our attempts using the second approach, to be detailed in the next paragraph. The graphic interface of {\tt SOLEX12} \citep{1997CeMDA..66..293V} was utilized to conveniently monitor the backward orbital evolution of the clones of the components, in which the orthogonal Cartesian coordinates of the MC clones of the fragments with respect to the position of the nominal orbit of the primary referenced to the J2000 ecliptic were plotted. We found  that around 2011 June, the MC clones of Fragment A formed a highly elongated ellipsoid spatially overlapping with that of the primary body, while the overlap between the MC clones of Fragment B appear to occur in a wider time range between early- and mid- 2011, due to the larger orbital uncertainty. However, the clones of Fragment C and those of the primary show no more than a partial overlap even during their closest encounter in mid- to late-2011. On the contrary, its MC clones overlap better with those of Fragment B from 2011 to 2014. This may imply that Fragments A and B separated  from the primary, whereas Fragment C possibly split instead from Fragment B. (Another possibility is that a nongravitational acceleration could exist, and so we consider this possibility in the second approach, below.) However, these results from the first approach are highly ambiguous, because the orbital uncertainties in backward integrations (dominated by the systematic errors in the astrometric reduction of field sources in the HST observations) are  enormous. Nevertheless, it is interesting that the derived separation times are broadly consistent with the epoch of  impulsive activity deduced independently from models of the tail morphology by \citet{2012ApJ...761L..12M} and \citet{2012ApJ...759..142S}. If we monitored the MC clones further backward in time, we would then observe seemingly acceptable overlaps between the MC clones of the primary and those of Fragments B and C in 2006 again. However, this is due to the orbital periodicity and the growing uncertainty regions with time backwards, and therefore we did not consider possible fragmentation epochs much earlier than 2011.

Our second approach is an application of the fragmentation model first devised by \citet{1977Icar...30..574S,1978Icar...33..173S}. We have previously succeeded in applying this approach to examine the split event of another active asteroid, P/2016 J1 (PANSTARRS) \citep{2017AJ....153..141H}. The model simplifies a split event as an instantaneous separation of one component from the other at a fragmentation epoch of $t_{\rm frg}$. At the instant after the split, a component travels at a separation velocity with respect to the other one referenced to the orbital plane of the latter and expressed in terms of the radial, transverse, and normal (RTN) components ($\Delta V_{\rm R}$, $\Delta V_{\rm T}$, and $\Delta V_{\rm N}$). If no satisfactory gravity-only solution is found, the RTN nongravitational parameters \citep[denoted as $A_1$, $A_2$, and $A_3$, respectively;][]{1973AJ.....78..211M}, are added to the model as additional free parameters to account for a potential nongravitational effect arising from any anisotropic recoil force acting on the fragment. The gravitational interaction between the split components is ignored, but the influence from the Sun as well as massive perturbers such as major planets in DE440 are taken into account. Then, the trajectory and apparent positions of the fragment can be uniquely determined by the above set of split parameters, and can be compared to the astrometric observations. Our code utilizes the Levenberg-Marquardt optimization code {\tt MPFIT} \citep{2009ASPC..411..251M} to obtain the split parameters that minimise the goodness of fit,

\begin{equation}
\chi^{2} \left(t_{\rm frg}, \Delta V_{\rm R}, \Delta V_{\rm T}, \Delta V_{\rm N}, A_1, A_2, A_3 \right) = {\bm \xi}^{\rm T} {\bf W} {\bm \xi}
\label{eq_chi2},
\end{equation}

\noindent where ${\bm \xi}$ is the vector of the O-C astrometric residuals, and ${\bf W}$ is the diagonal weight matrix of the observations assigned as the inverse square of the astrometric measurement uncertainties. In order to eliminate systematic uncertainties in our absolute astrometric measurements, we fitted the relative astrometry of the paired components. In this way the associated uncertainties arise primarily from errors in the best-fit centroids of the components. These are typically smaller than the overall errors listed in Table \ref{tab_astrom} by at least an order of magnitude.

We found satisfactory solutions without the  need to include non-gravitational acceleration terms (the RTN nongravitational parameters were all set to be zero). Given also the statistics of separation speeds of comets \citep[and citations therein]{2004come.book..301B} and the estimate on the ejection speed of dust based on modelling the dust trail morphology \citep{2021AJ....162..268J}, we confined the magnitudes of the RTN components of the separation velocity to be $\le\!10$ m s$^{-1}$. The Levenberg-Marquardt optimization is only capable of finding local minima, and the fragmentation epoch is the only parameter that could vary substantially. Therefore, we first  examined how the goodness of fit changes as a function of the fragmentation epoch by iteratively solving, for the three fragments, their separation velocities with the fragmentation epoch fixed in a window $\pm$6 month from the  impulsive activation epoch of \citep[2011 July;][]{2012ApJ...761L..12M,2012ApJ...759..142S}. Once each iteration was completed, we recorded the best-fit separation velocity and the normalized root-mean-square residual of the fit, which is the goodness of fit divided by twice the number of observations (so as to get rid of the difference in the numbers of observations between fragments), then incremented the fragmentation epoch by a step size of one day, and started a new fit by repeating the same  procedures.

\begin{figure*}
\epsscale{1.0}
\begin{center}
\plotone{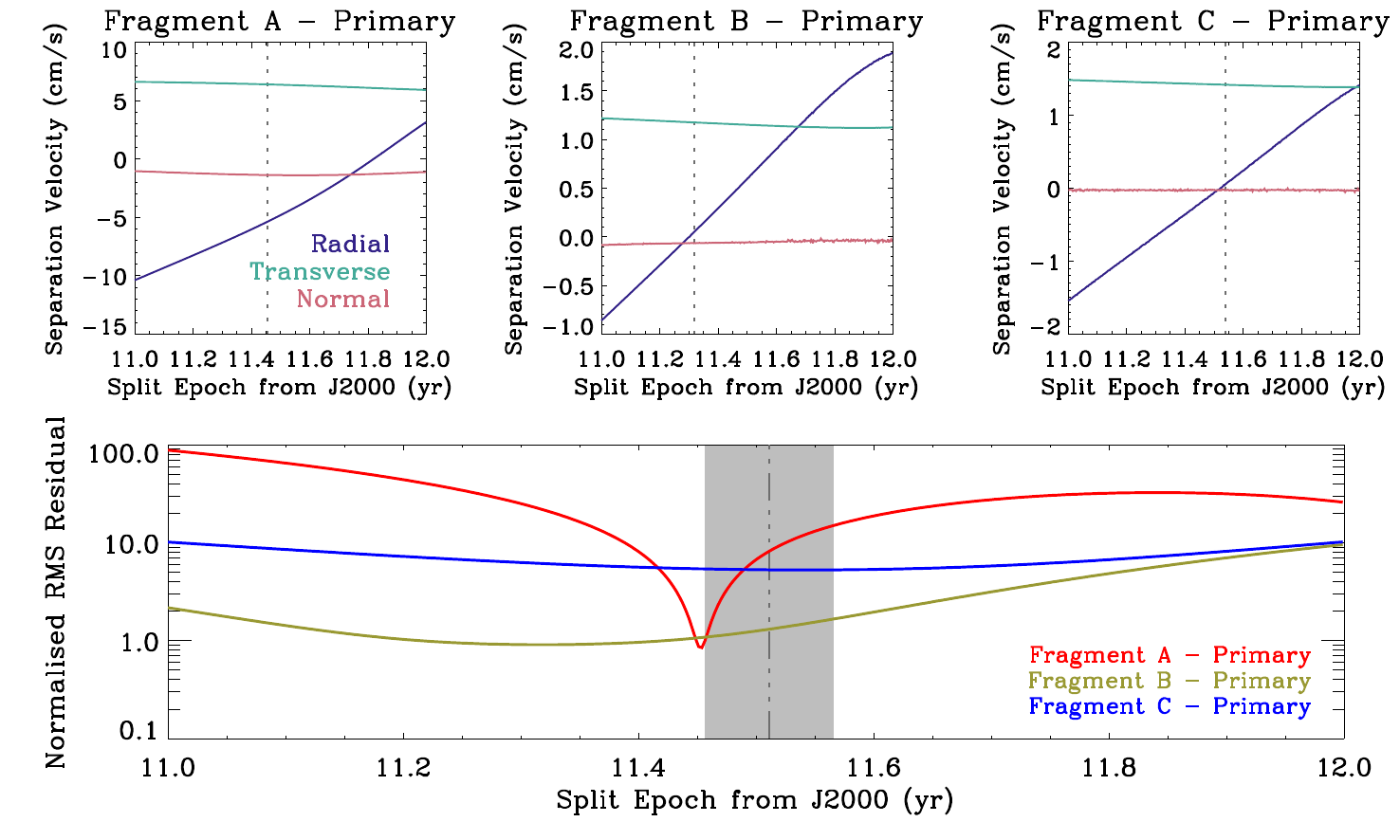}
\caption{
Upper three panels: separation velocities decomposed into radial, transverse, and normal components (color coded) of Fragments A (left), B (middle), and C (right) split from the primary of 331P needed in order to best fit the astrometry as functions of fragmentation epoch (here expressed as number of years past J2000). Lower panel: normalized rms residuals (the goodness of fit divided by twice the number of observations) of the best fits for Fragments A, B, and C (distinguished by colors) varied with split epoch. The shaded region represents the range of possible impulsive activation epochs of 331P inferred from the tail morphology by \citet[2011 July $7 \pm 20$]{2012ApJ...759..142S}, with the nominal value marked by the vertical line. The dotted lines in grey in the upper panels indicate split epochs rendering solutions with corresponding local minimum normalized rms residual. Note the best normalized rms for Fragment C is worse than those for the other two fragments by an order of magnitude.
\label{fig:search_tfrg1}
} 
\end{center} 
\end{figure*}

The upper panels of Figure \ref{fig:search_tfrg1} display the resulting relative separation velocities while the lower panel compares the normalized rms residuals of the three fragments. In the 2011 - 2012 timeframe, there is only a single minimum in the normalized rms residuals for each of the fragments. Two major differences can be readily noticed. First, the minimum of the normalized rms residual of Fragment A is narrow and much better defined than those of Fragments B and C.  The Fragment A minimum occurs at $\sim$2011.45 (2011 June 16), consistent with the activation dates 2011 July $7 \pm 20$ inferred from tail measurements  \citep{2012ApJ...759..142S}.  The Fragment B minimum occurs near 2011.3 but, given its width, is consistent with that of Fragment A and with \cite{2012ApJ...759..142S}.  Second,  the minima in the normalized rms residuals of Fragments A and B are both $\la \! 1$, representing plausible solutions but, in contrast, the minimum in the computed separation of  Fragment C from the primary has a normalized rms $\sim$10. All measurements of the 331P to Fragment C separation  from 2015 show O-C residuals more than ten times larger than the measurement uncertainties, and nearly all observations from 2017 show residuals exceeding the $3\sigma$ level. We conclude that while Fragments A and B could both have separated from the primary body and have done so simultaneously,  in contrast, the solution for the separation of Fragment C  from the 331P primary body is  unacceptable.

\begin{figure*}
\epsscale{1.0}
\begin{center}
\plotone{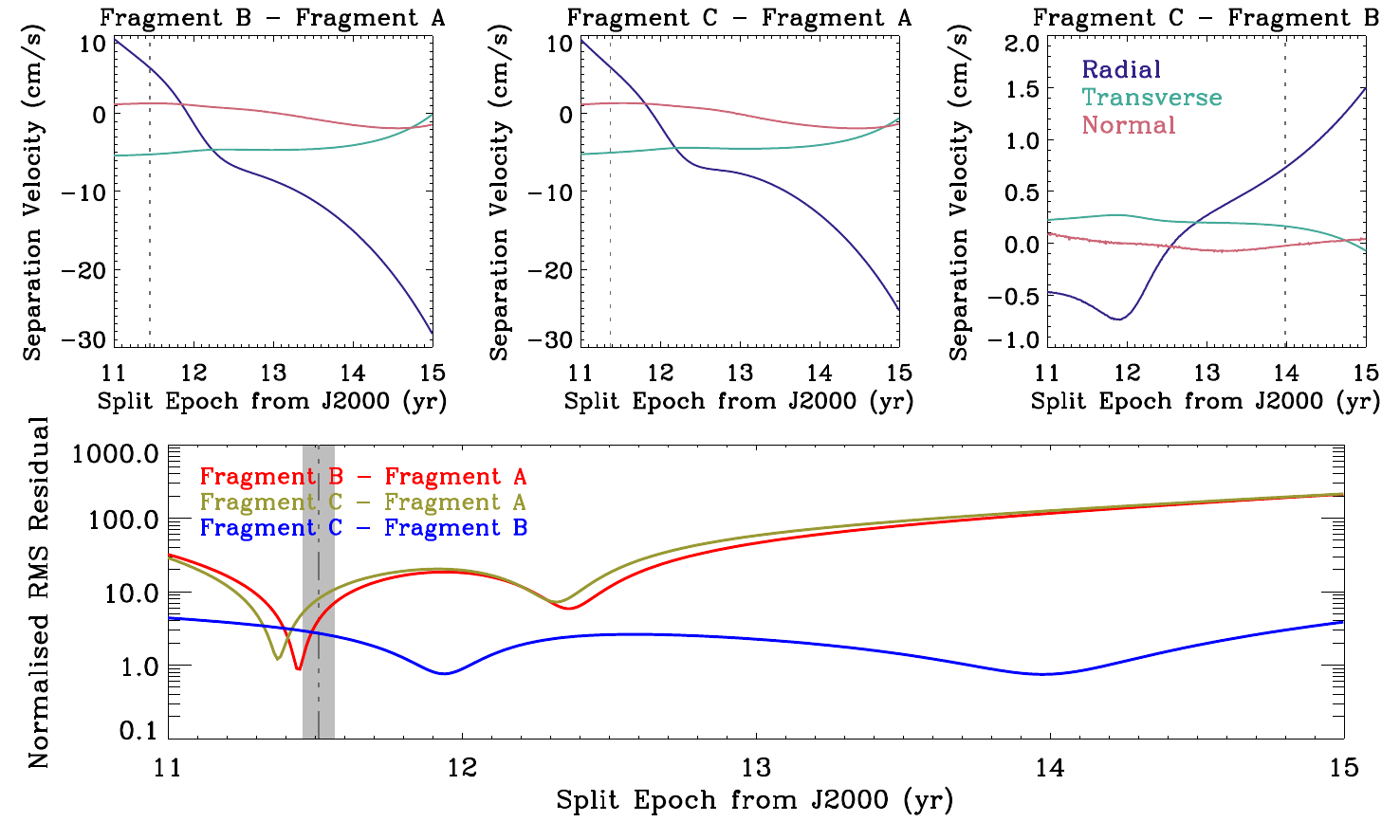}
\caption{
Same as Figure \ref{fig:search_tfrg1} but for pairwise combinations between the fragments themselves. The normalized rms residuals all have two local minima in the investigated split epoch range, but only in the pair of Fragments B and C, are the local minima comparable to each other, as well as to those for Fragments A and B in Figure \ref{fig:search_tfrg1}.
\label{fig:search_tfrg2}
} 
\end{center} 
\end{figure*}

There are three additional pairwise combinations of the fragments to be investigated, i.e., Fragments A and B, Fragments A and C, and Fragments B and C. Given the ambiguous results we observed in the backward integration of the MC clones, we examined the possible split epochs in a range still from early 2011 extending to early 2015, almost a year before the earliest HST observation. The above  fitting procedures were repeated, except that the step size was increased to four days so as to reduce the computation loads. We present the results in Figure \ref{fig:search_tfrg2}, in which we can see that there are two local minima in each of the normalized rms residuals of the solutions for the three pairwise combinations. However, only the pairwise consideration of Fragments B and C has two local minima of comparable depth, with normalized rms $\sim\!1$. In comparison, the best fit for the separation between Fragments A and C is worse, in that one observation has an O-C residual of nearly $4\sigma$ and one of almost $3\sigma$, both in the declination direction. The weighted mean residual in the declination direction is more than twice the one in the J2000 equatorial east-west direction, which is absent in any other best solutions. More importantly, the best-fit fragmentation epoch (around 2011 May 19) would predate the split events that produced Fragments A and C. Therefore, we conclude that the split between Fragments A and C is implausible and the hypothesis is rejected.

Having known the numbers of local minima in the normalized rms residuals in the time interval, we relaxed the fragmentation epoch as a free parameter and tested with different combination sets of initial guess values for the pairwise combinations of the components of 331P. We found  that each best-fit solution would always converge to the same results within the uncertainty levels, as long as the initial guess values for the split epoch is situated within the dip of the normalized rms residual. The individual O-C astrometric residuals of the best-fit solutions for the pairwise combinations of Fragment A and the primary, Fragment B and the primary, Fragments B and A, and Fragments C and B are all consistent with the measured uncertainties. Table \ref{tab_split} tabulates these best-fit solutions, the reported uncertainties are all formal $1\sigma$ errors properly propagated from the astrometric measurement errors.

We were fully aware that the derived uncertainties of the split parameters have not incorporated the orbital uncertainties of the paired components. To investigate how strong the influence of the latter would be, we carried out iterative runs with the orbits of the referenced components substituted by the corresponding MC clones. In each run, we obtained the best-fitted fragmentation epoch and the RTN components of the separation velocity and recorded the corresponding goodness of fit and normalized rms residual by fitting the differential astrometry of the fragments. After the whole run was completed, we then computed the standard deviations of the split parameters, finding that even in the cases with the least certain orbital solutions (for Fragments B and C), the resulting uncertainties are always less than a hundredth, typically less than a thousandth, of the counterparts propagated from the measurement errors alone. Therefore, we believe that the reported uncertainties in Table \ref{tab_split} are trustworthy.

\begin{deluxetable*}{lCccccc}
\tablecaption{Plausible Gravity-only Solutions to the Split Parameters of Fragments A, B, \& C of Active Asteroid 331P/Gibbs
\label{tab_split}}
\tablewidth{0pt}
\tablehead{
\multicolumn{2}{c}{Quantity}  & 
\colhead{Fragment A} &
\multicolumn{2}{c}{Fragment B} &
\multicolumn{2}{c}{Fragment C} \\
\cmidrule(lr){4-5} \cmidrule(lr){6-7}
&&&
\colhead{Solution I} &
\colhead{Solution II} &
\colhead{Solution I} &
\colhead{Solution II}
}
\startdata
Fragmentation epoch (TDB)\tablenotemark{$\dagger$} & t_{\rm frg}
       & 2011 Jun $15.62 \pm 0.32$
       & 2011 Apr $26 \pm 26$
       & 2011 Jun $12.15 \pm 0.83$
       & 2011 Dec $10 \pm 12$
       & 2013 Dec $23 \pm 18$ \\ 
Separation velocity (cm s$^{-1}$) &&&&\\
\hspace{5mm}Radial component & \Delta V_{\rm R}
       & $-5.398 \pm 0.012$
       & $+0.04 \pm 0.24$
       & $+5.939 \pm 0.026$
       & $-0.723 \pm 0.045$
       & $+0.724 \pm 0.030$ \\
\hspace{5mm}Transverse component & \Delta V_{\rm T}
       & $+6.39473 \pm 0.00095$
       & $+1.1767 \pm 0.0082$
       & $-5.2422 \pm 0.0024$
       & $+0.2715 \pm 0.0050$
       & $+0.1661 \pm 0.0045$ \\ 
\hspace{5mm}Normal component & \Delta V_{\rm N}
       & $-1.3693 \pm 0.0094$
       & $-0.064 \pm 0.023$
       & $+1.309 \pm 0.027$
       & $-0.001 \pm 0.028$
       & $-0.024 \pm 0.033$ \\ 
\hline
\multicolumn{2}{l}{Paired with}
& Primary
& Primary
& Fragment A
& Fragment B
& Fragment B \\
\multicolumn{2}{l}{Normaliznormalizeed rms residual}
& 0.834
& 0.910
& 0.826
& 0.761
& 0.750 \\
\enddata
\tablenotetext{\dagger}{The corresponding uncertainties are in days.}
\tablecomments{Our best-fit gravity-only solution for Fragment C in the scenario where it is assumed to split from the primary renders an unacceptably strong systematic trend in the O-C astrometric residuals, thereby not presented here. The reported uncertainties are $1\sigma$ formal errors. The separation velocities are referenced with respect to the corresponding paired components of 331P.}
\end{deluxetable*}

\section{Discussion}
\label{sec_dsc}

\begin{figure*}
\epsscale{1.0}
\begin{center}
\plotone{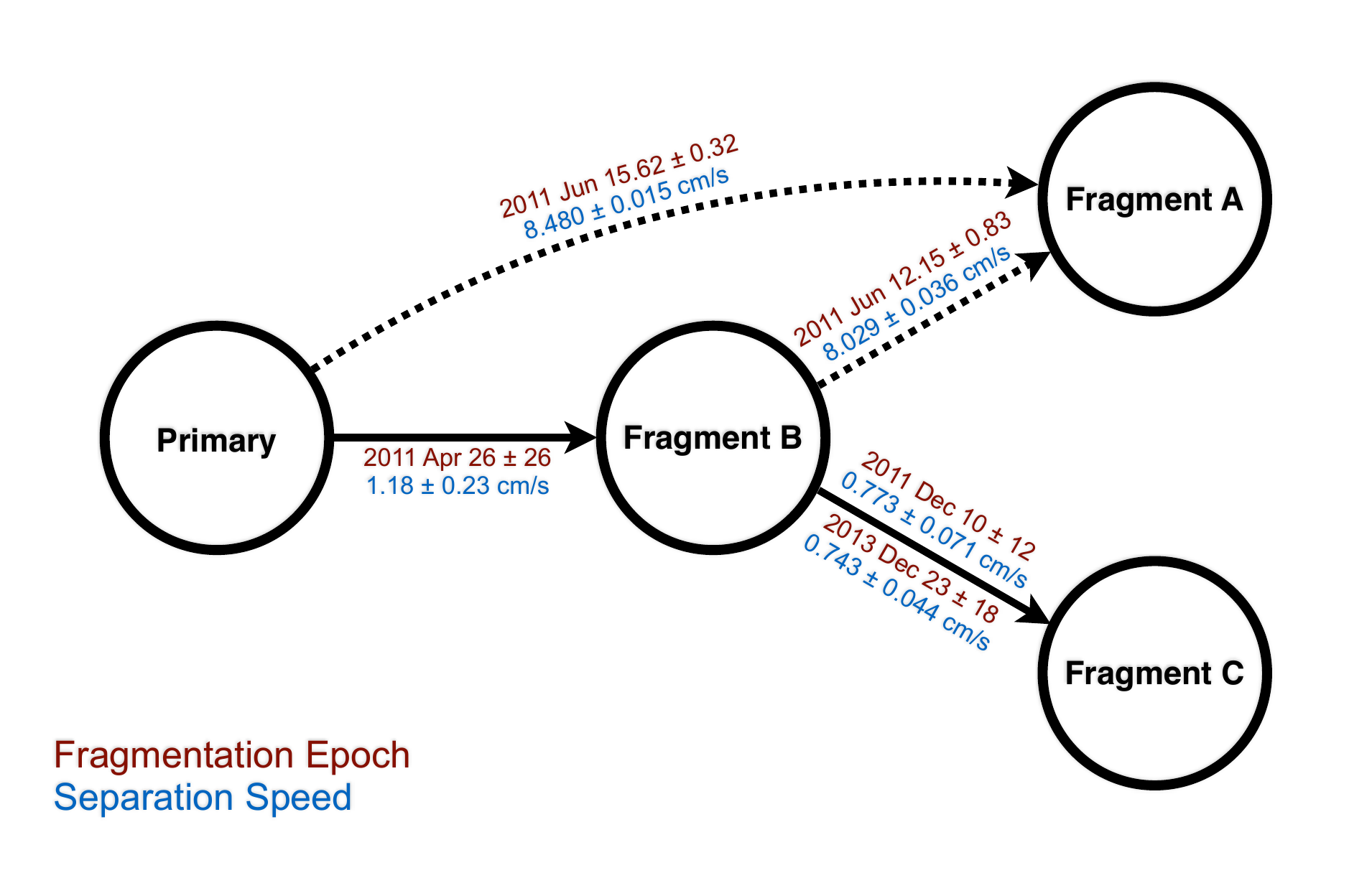}
\caption{
Schematic diagram showing plausible relations between the four brightest  components of 331P. Texts in dark red and blue respectively label the fragmentation epoch and the separation speed with respect to the pairwise parent components. Dotted arrows indicate the existence of more than a single possible pairwise relationship. Two plausible split scenarios were found for the pair of Fragments B and C, with the major difference in the split epoch.
\label{fig:frg_scheme}
} 
\end{center} 
\end{figure*}

One may ask whether or not there can be other fits to the data if nongravitational accelerations are  included in the model, and we explored this possibility. Given the fact that the in-plane components of the nongravitational acceleration are usually orders of magnitude larger than the out-of-plane one, we only tested with nonzero $A_1$ and/or $A_2$, while setting $A_3 \equiv 0$ au day$^{-2}$ in all cases.   We find that acceptable solutions including non-gravitational acceleration parameters are possible.  For example, we tested the separation of Fragment C from the primary with the inclusion of a radial non-gravitational acceleration, $A_1$. We found a solution with  goodness of fit even better than any of the gravity-only solutions, having a normalized rms residual of 0.683, provided $A_1 = \left(2.73 \pm 0.45 \right) \times 10^{-10}$ au day$^{-2}$. Another formally acceptable fit was obtained by instead assuming $A_1 = A_3 = 0$ au day$^{-2}$, giving a normalized rms residual of only 0.791. However, the apparent improvement of these  solutions is almost certainly an artifact of adding more parameters to the fitting model.  

Physical measurements of the 331P fragments  show no evidence for comet-like outgassing as would be expected in the presence of sublimating ice  \citep{2021AJ....162..268J}.  In the absence of sublimation, nongravitational accelerations can occur from  solar radiation pressure and the Yarkovsky effect, but these are very small. The corresponding radial nongravitational parameter is

\begin{equation}
A_1 = \frac{3 \left(1 + A_{\rm B} \right) S_{\odot}}{4 c \rho R}
\label{eq_A1},
\end{equation}

\noindent where $A_{\rm B}$ is the Bond albedo, $S_{\odot} = 1361$ W m$^{-2}$ is the solar constant, $c \approx 3 \times 10^8$ m s$^{-1}$ is the speed of light in vacuum, and $\rho$ and $R$ are respectively the mass density and the radius of the fragment. Substituting with the size and mass density estimates \citep{2021AJ....162..268J} and the Bond albedo $A_{\rm B} \sim 0$, we find the anticipated radial nongravitational parameter to be $A_1 \sim 10^{-12}$ au day$^{-2}$ for Fragment C. This is two orders of magnitude smaller than the acceleration deduced from the model. We used JPL's Small-Body Database Query\footnote{\url{https://ssd.jpl.nasa.gov/tools/sbdb_query.html}} to search for  comparably small asteroids with measured $A_2$ values. These small bodies all have $\left|A_2 \right| \sim 10^{-13}$ au day$^{-2}$, again  two orders of magnitude smaller than the value returned by our model. It is unsurprising to find apparently better solutions  by  adding extra free parameters to the model, but these solutions have no physical significance and gravity-free solutions are just as good.

Given the obtained best-fit fragmentation epochs and the associated uncertainties, we can think of two different possibilities for the fragmentation events that occurred at 331P:

\begin{enumerate}
\item Fragments A and B  both directly separated from the primary in 2011 mid-June at a separation speed of $\sim$8 cm s$^{-1}$ and between April and May 2011 at $\sim$1 cm s$^{-1}$, respectively, while Fragment C split from Fragment B at a later epoch, either around late 2011 or between late 2013 and early 2014, at a speed of $\sim \! 0.7$-0.8 cm s$^{-1}$.

\item Fragment B is the only component that directly fragmented from the primary around April and May in 2011, and it subsequently released Fragment A in 2011 mid-June with a separation speed of $\sim$8 cm s$^{-1}$ and then Fragment C, either in late 2011, or between late 2013 and early 2014.

\end{enumerate}

\noindent We summarize the two possibilities in Figure \ref{fig:frg_scheme}, in which the separation speeds between the components are labelled. The listed uncertainties are formal $1\sigma$ errors calculated using the covariance matrices of the split parameters. Unfortunately, other fragments of 331P visible in the HST observations do not have long enough observing arcs, and therefore their fragmentation scenarios cannot be meaningfully constrained. Admittedly, a drawback of our fragmentation model is that mutual gravitational influences between the components are ignored. However, as the mass, shape, and density distribution will all strongly modify the gravitational potential of a component, all of which are unknown, and treating the components of 331P as multiple massive bodies will almost certainly give rise to chaos in the system of 331P. Given these, in addition to the measurement uncertainties that would otherwise leave too much room for various possible parameter options, we do not see any benefit in taking these extra factors into consideration in the fragmentation model, and we believe that the our results are diagnostic enough. The component separations at different epochs strongly imply that 331P is  undergoing cascading fragmentation. Its components form an asteroid cluster that, at just over a decade, is considerably younger than any other known asteroid cluster \citep[and citations therein]{2022arXiv220506340N}. Cascading fragmentation has been previously identified in other asteroid clusters \citep{2020Icar..33813554F}, including the active asteroid P/2013 R3 (see Figure 12 of \cite{Jewitt17}). Therefore, the ongoing fragmentation at 331P provides us with a great opportunity to study and understand fragmentation of asteroids.

Using the derived separation velocities and times,  we qualitatively explore the most likely physical mechanism that led to the cascading fragmentation at 331P. The best-fit separation velocities lie close to the orbital plane of 331P, as the in-plane components were found to be at least around five times larger than the out-of-plane components. The fragmentation speeds between the components are in the range  $\sim$0.7-8 cm s$^{-1}$. As per the size estimates of the primary and the three largest fragments by \citet{2021AJ....162..268J}, the minimum ejection speeds required to escape the gravitational potential of the primary and that of Fragments A, B, and C are $\sim \! 0.7$ m s$^{-1}$ and $\sim \! 6$-10 cm s$^{-1}$, respectively. Since our fragmentation model neglects the mutual gravitational forces between the components, the measured separation speeds should be seen as the residual speeds at infinity with respect to the pairwise component. Assuming the components are all spherical, we roughly estimate the ejection speed to be

\begin{equation}
V_{\rm ej} = \sqrt{\Delta V^2 + \frac{8}{3} \pi \rho G \left(R_{1}^2 - R_{1} R_{2} + R_{2}^2 \right)}.
\label{eq_Vej}
\end{equation}

\noindent Here, $G = 6.67 \times 10^{-11}$ m$^{3}$ kg$^{-1}$ s$^{-2}$ is the gravitational constant, and $R_1$ and $R_{2}$ are the nucleus radii of the pairwise components. Substituting, we find $V_{\rm ej} \approx 0.7$ m s$^{-1}$ for Fragments A and B to split from the primary, 0.1 m s$^{-1}$ for the split between Fragments A and B, and 7 cm s$^{-1}$ for the split between Fragments B and C, all of which are comparable to the respective escape speeds.  

The best-fit results from our astrometric model of fragmentation  indicate that the split events were spread over several months, inconsistent with an impact origin. In the absence of other plausible mechanisms, we concur with \citet{2015ApJ...802L...8D} and \citet{2021AJ....162..268J} that the fragmentation at the primary of 331P was most likely caused by rotational instability.  Separation speeds below the nominal escape velocity, as inferred from the data, are natural products of multi-body interactions in a rotationally fissioned asteroid, where delayed ejection of material is also a prominent feature of the models.  For example, numerical simulations of rotationally fissioned asteroids reveal complex interactions that produce cm s$^{-1}$ excess escape velocities  in abundance, commonly result in secondary fission, and can have survival lifetimes measured in years \citep{Boldrin16}.    

\begin{figure}
\epsscale{0.75}
\begin{center}
\plotone{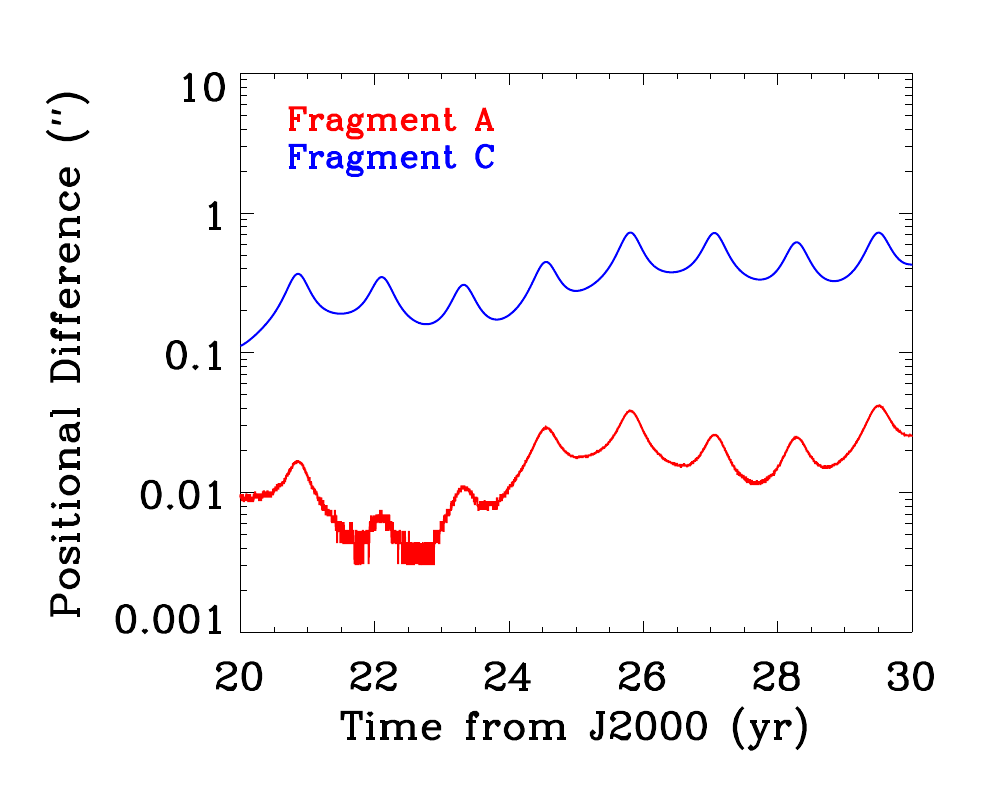}
\caption{
Ephemeris positional differences between the possible fragmentation scenarios of Fragments A (red) and C (blue) from 2020 to 2030. The sawtooth features in the curve of Fragment A are artifacts due to the fact that the positional difference approaches the output precision limit of the computed J2000 equatorial coordinates in the ephemerides.
\label{fig:ephem_dpos}
} 
\end{center} 
\end{figure}

Future observations at HST resolution and sensitivity are needed to resolve the ambiguities of the present work and will reveal the dynamics of this evolving system more clearly.  To explore this, we integrated the motions of the fragments into the future. We find that by 2023, the two possible fragmentation scenarios described for Fragment C, above, will lead to a positional difference of $\sim \! 0\farcs3$ (see Figure \ref{fig:ephem_dpos}).  This is large enough that new measurements from HST (or JWST) will be able to distinguish between the scenarios.   New observations of comparable or greater sensitivity may also show the spread of other fragments, and reveal whether the fragmentation cascade continues.

\section{Summary}
\label{sec_sum}

We present an astrometric analysis of the multi-component active asteroid 331P/Gibbs using data obtained from the Hubble Space Telescope from 2015 to 2018.  We applied a fragmentation model to  Fragments A, B, and C which, together with the primary body, 331P, are the brightest and best observed components.  The key results are:

\begin{enumerate}
\item Fragment B split from the primary  in 2011 April to May, with a separation speed of $1.2 \pm 0.2$ cm s$^{-1}$.  This date is in broad agreement with initiation dates determined independently from models of the particle trail in which the fragments appear embedded.

\item We find two plausible solutions for Fragment A, with separation from the primary on 2011 June $15.62 \pm 0.32$ at a relative speed of $8.48 \pm 0.02$ cm s$^{-1}$, or  from Fragment B slightly earlier on 2011 June $12.15 \pm 0.83$ at speed $8.03 \pm 0.04$ cm s$^{-1}$. 

\item We find no acceptable solutions in which Fragment C is released directly from the primary body.  Instead, we  find two possible scenarios in which Fragment C split from Fragment B, either  in late 2011 at  $0.77 \pm 0.18$ cm s$^{-1}$ or near late 2013 to early 2014 at  $0.74 \pm 0.04$ cm s$^{-1}$.

\item The small fragment separation speeds and the staggered epochs of separation are most compatible with rotational disruption as the mechanism that led to the breakup of 331P starting in 2011.

\end{enumerate}

\begin{acknowledgements}
We thank the anonymous reviewer for insightful and helpful suggestions and comments. The codes we employed to perform absolute astrometric measurements were authored and  shared by David J. Tholen. We exploited {\tt AstroMagic} by Gianpaolo Pizzetti and Marco Micheli for visualization and extraction of star trails in the HST images. We thank Bill Gray and Aldo Vitagliano respectively for implementing {\tt FindOrb} and {\tt SOLEX12} as per our fastidious requests and making them publicly available. JPL's Horizons API was extensively exploited in this work.  Based on observations made with the NASA/ESA Hubble Space Telescope, obtained from the data archive at the Space Telescope Science Institute. STScI is operated by the Association of Universities for Research in Astronomy, Inc. under NASA contract NAS 5-26555. Support for this work was provided by NASA through grant number AR-16618 from the Space Telescope Science Institute, which is operated by AURA, Inc., under NASA contract NAS 5-26555.
\end{acknowledgements}

\vspace{5mm}
\facilities{HST/WFC3}

\software{{\tt AstroDrizzle} \citep{2012drzp.book.....G}, {\tt FindOrb}, {\tt IDL}, {\tt L.A. Cosmic} \citep{2001PASP..113.1420V}, {\tt MPFIT} \citep{2009ASPC..411..251M}, {\tt SOLEX12} \citep{1997CeMDA..66..293V}}

\end{document}